\documentclass[twocolumn,showpacs,showkeys]{revtex4}
\usepackage{graphicx}
\usepackage{bm}
\usepackage{color}
\usepackage{amsmath}
\usepackage{natbib}

\begin{document}

\title{Destabilization effect of exchange dipole-dipole interaction on the spectrum of electric dipolar ultracold Fermi gas}

\author{Pavel A. Andreev}
\email{andreevpa@physics.msu.ru}
\affiliation{M. V. Lomonosov Moscow State University, Moscow, Russia.}

\date{\today}

\begin{abstract}
The self-consistent field approach for the electric dipolar ultracold spin-1/2 fermions is discussed. Contribution of the exchange part of the electric dipole interaction is found. Hence we obtain a model of dipolar fermions beyond the self-consistent field approximation. It is shown that the exchange interaction of electric dipolar fermions depends on the spin-polarisation of the system. For instance the electric dipole exchange interaction equals to zero for spin-unpolarised systems, namely all low laying quantum states occupied by two-particles with opposite spins. In opposite limit of the full spin polarisation of the degenerate fermions, then we have one particle in each quantum states, the exchange interaction has maximum value, which is comparable with the self-consistent field part of the dipole-dipole interaction. The self-consistent part of the electric dipole-dipole interaction gives a positive contribution into the spectrum of collective excitations, while the exchange part of the dipole-dipole interaction leads to a negative term in the spectrum. At angles between the equilibrium polarisation and the direction of wave propagation close to $\pi/2$ the full dipolar part of the spectrum becomes negative. At the electric dipole moment of fermions of order of several Debay the dipolar part is large enough to exceed the Fermi pressure, that reveals in an instability. We also consider spectrum of the quasi two-dimensional cloud of fermions in the trap. We consider the regime of purely two dimensional structure of dipolar fermions with the exchange dipole-dipole interaction in the three dimensional space and calculate the spectrum in this regime. We assume that the equilibrium polarisation is perpendicular to the two dimensional structure. Major picture of the spectrum behavior in low dimensional regimes is similar to the three-dimensional one. Since the two-dimensional perturbations propagate perpendicular to the equilibrium polarisation we find that the dipolar part of the spectrum is negative in these regimes.
\end{abstract}


\maketitle


\section{Introduction}

Dipolar Fermi molecules \cite{Ni Science 08}-\cite{Heo PRA 12} attract as mach attention as electrically polarised molecules in the Bose-Einstein condensate (BEC) state \cite{Deiglmayr PRL 08}-\cite{Munchow PCCP 11}. Majority of researches in the field of dipolar quantum gases are focused on the aligned dipoles \cite{Lahaye RPP 09}-\cite{Baranov CR 12}, but there are paper where evolution of the dipole directions is included as well \cite{Andreev EPJ D 13}-\cite{Andreev RPJ 13}.

A lot of achievements in physics of ultracold fermions were reviewed in Ref. \cite{Giorgini RMP 08}. The non-linear Schrodinger equation and the hydrodynamic equations have been applied for description of ultracold fermions \cite{Belemuk PRA 07}-\cite{Karpiuk PRA06}. Hydrodynamic model of dipolar ultracold fermions was derived in approximation of aligned dipoles applying the microscopic density matrix \cite{Lima PRA 10}. This model is in agreement with the similar model of dipolar BECs suggested earlier \cite{Santos PRL 00}, \cite{Yi PRA 00}, \cite{Goral PRA 00}, \cite{Yi PRA 01}. Collective excitations of quasi-two-dimensional trapped dipolar fermions were considered in Ref. \cite{Babadi_Demler PRA 12}. Different generalizations of the model of dipolar BECs were suggested. Contribution of the quantum fluctuations in the characteristics of dipolar BECs was considered in Ref. \cite{Lima PRA 12}. The dipole-dipole interaction, in terms of collisions, in the dipolar BECs, beyond the first Bohm approximation was developed in Ref. \cite{Wang NJP 08}. Spectrums of dipolar BECs at finite temperatures were calculated in Refs. \cite{Ticknor PRA 12}, \cite{Natu_Wilson PRA 13}. Exchange effects in temperature distributed quantum gases (bosons and fermions) were considered in Refs. \cite{Baillie PRA 12}, \cite{Baillie PRA 12 (R)}. Non-integral Gross-Pitaevskii equations for dipolar (electric and magnetic) BECs were obtained in Refs. \cite{Andreev MPL 13}, \cite{Andreev EPJ D 14}. Difference between the magnetic and electric dipolar BECs was discussed in Refs. \cite{Andreev EPJ D 13}, \cite{Andreev EPJ D 14}. Dipole direction evolution and its influence on spectrum of electric dipolar BECs were described in Refs. \cite{Andreev EPJ D 13}, \cite{Andreev RPJ 12}. Model of ultracold dipolar fermions  with the dipole direction evolution was developed in Refs. \cite{Andreev RPJ 13}, \cite{Andreev LongetDipFermi12}.

Fisher theoretically demostrated stability of quasi two-dimensional dipolar Bose-Einstein condensates \cite{Fischer PRA 06R}. This result extended the list of fundamental results on the trapped dipolar Bose-Einstein condensates \cite{Goral PRA 02}-\cite{Wilson PRL 08}. Further analysis of general properties of the flattened dipolar condensates was presented in Ref. \cite{Baillie_arx 14}. The evolution of correlations in a quasi-two-dimensional dipolar gas driven out of equilibrium by a sudden ramp of the interactions was investigated in Ref. \cite{Natu+Sarma PRA 14}. Anisotropic superfluidity in dipolar Bose-Einstein condensates in a quasi-two-dimensional geometry was considered \cite{Ticknor PRL 11}. Fedorov et al. \cite{Fedorov PRA 14} predicted the effect of the roton instability for a two-dimensional weakly interacting gas of tilted dipoles in a
single homogeneous quantum layer. Low dimensional dipolar fermions \cite{Lima PRA 10}, \cite{Sieberer ar 11} attract a lot of attention along with the quasi two-dimensional dipolar Bose-Einstein condensates. The strong correlations on the phase diagram and collective modes of
quasi-two-dimensional dipolar fermions were considered in Ref. \cite{Babadi arx 12}. The density-wave phase
of a two-dimensional dipolar fermions was studied in Refs. \cite{Marchetti arx 12}, \cite{Block Bruun arx 14}. The BCS superfluid transition in a single-component fermionic gas of dipolar particles
loaded in a tight bilayer trap was studied in Ref. \cite{Baranov arx 10}. The three-dimensional dipolar fermions \cite{Liu Li Yin}, \cite{Adhikari ar 12 fermi}, two component systems of dipolar bosons \cite{Wilson PRA 12}, and dipolar boson-fermion mixtures \cite{Adhikari arx 13 mixt} are also under consideration. 
Jona-Lasinio et al demonstrated that the density dependence of the roton minimum results in a spatial roton confinement, that roton confinement plays a crucial role in the dynamics after roton instability, and that arresting the instability may create a trapped roton gas
revealed by confined density modulations \cite{Jona-Lasinio 13}. The ground state of a system of bosons with aligned dipoles located in a plane is analyzed in Ref. \cite{Macia PRA 14}. The Feshbach resonances in dipolar quantum gases were described by Kotochigova \cite{Kotochigova arx 14}.

In this paper we consider the fermi molecules with the aligned electric dipoles. We use the method of many-particle quantum hydrodynamics to derive the model. We consider dipolar fermions beyond the self-consistent field approximation, so we calculate contribution of the exchange electric dipole-dipole interaction. We obtain that the exchange part of the interaction is comparable with the self-consistent field part for fully spin polarised spin-1/2 fermions possessing the electric dipole moment. Contribution of the exchange interaction becomes smaller with decreasing of the spin polarisation. It equals to zero for the spin unpolarised systems of fermions with the aligned electric dipoles. The dynamic of spins and the spin-spin interaction are not considered in this paper, since we focused on the electric dipole contribution, and the electric dipoles have rather larger contribution than the contribution of the magnetic dipoles. However we mention spin of particles, since their spin states give influence on the exchange part of the electric dipole-dipole interaction.
We should note that the exchange spin-spin interaction in systems of spin-1/2 particles was derived in Ref. \cite{MaksimovTMP 2001 b}.

The self-consistent field part and the exchange part of the electric dipole interaction have different signs. Thus the exchange electric dipole interaction can cause an instability of the three dimensional cloud of dipolar fermions. Therefore we have calculated the spectrum of collective excitations (an analog of the Bogoliubov spectrum of dipolar Bose-Einstein condensates) of the quasi-two-dimensional and quasi-one-dimensional Fermi gas. We also calculate spectrums for two-dimensional and one-dimensional cases to compare it with the quasi-two-dimensional and quasi-one-dimensional systems.

Dipolar Bose-Einstein condensates (BECs) in terms of the quantum hydrodynamic method beyond the self-consistent field approximation was considered in Ref. \cite{Andreev Exchange_DD_BEC}. This consideration revealed a strange result. It was shown that the self-consistent field part equals to zero. Hence all the dipole-dipole interaction in dipolar BECs is related to the exchange dipole-dipole interaction. Moreover the explicit form of the exchange interaction of the electric dipoles coincides with the result of formal application of the self-consistent field approximation. It was shown that this conclusion is correct for both the aligned dipoles and systems of dipoles with the dipole-direction evolution. Formal application of the self-consistent field approximation in both described cases was performed in Refs. \cite{Andreev EPJ D 13}, \cite{Andreev RPJ 12}, \cite{Andreev JP B 14}, \cite{Andreev MPL 13}, \cite{Andreev EPJ D 14}. This consideration was made to give the microscopic derivation of equations governing the dynamics of dipolar BECs, which, for the case of aligned dipoles, was suggested in Refs. \cite{Santos PRL 00}, \cite{Yi PRA 00}, \cite{Goral PRA 00}. Methods developed in Refs. \cite{MaksimovTMP 2001 b}, \cite{MaksimovTMP 1999}-\cite{Andreev 1407 Review} were applied in Ref. \cite{Andreev Exchange_DD_BEC}.

Considering the exchange interaction in this paper we follow Refs. \cite{MaksimovTMP 2001 b}, \cite{MaksimovTMP 1999}, \cite{Andreev PRA08}, \cite{Andreev AoP 14}, \cite{Andreev 1407 Review}. While, there are a number fundamental papers on other methods of dealing with the exchange interaction, which were developed for three- and two-dimensional electron gas \cite{Nozieres PR 58}-\cite{Datta JAP 83}.
A kinetic model for finite temperature neutral boson and fermion gases including the exchange dipole-dipole interaction was developed in Refs. \cite{Baillie PRA 12}, \cite{Baillie PRA 12 (R)}.

Collective excitations of a harmonically trapped, two-dimensional, spin-polarized
dipolar Fermi gas in the hydrodynamic regime was considered in Ref. \cite{Zyl arXiv14}, where authors applied the Thomas-Fermi von Weizs\"{a}cker energy functional for the neutral atoms with the magnetic moments. They obtained the force field $\textbf{F}$ of the magnetic dipole-dipole interaction proportional to $\nabla n_{2D}^{5/2}$ (in their paper they have $\textbf{F}\sim \nabla n_{2D}^{3/2}$ due to different definition of the force $\textbf{F}$). Their formalism is based on the two-dimensional of particles $[n_{2D}]=cm^{-2}$. The structure of the full force of the magnetic dipole-dipole interaction obtained in Ref. \cite{Zyl arXiv14} corresponds to the exchange part of the exchange electric dipole interaction derived in our paper for purely two dimensional dipolar fermions (see formula (\ref{Bscf_Fer bal imp eq 2D}) below). Let us also note that quantum hydrodynamic derivation of the force of magnetic dipole-dipole interaction (the spin-spin interaction) for three dimensional systems of particles with the exchange spin-spin interaction was presented in Ref. \cite{MaksimovTMP 2001 b}.

This paper is organized as follows. In Sec. II we describe the model of dipolar degenerate fermions in the self-consistent field approximation for align electric dipoles. In Sec. III we present generalisation of the model including the exchange part of the electric dipole-dipole interaction. In Sec. IV  we consider excitation spectrum of three-dimensional plane waves. In Sec. V we consider quasi-two-dimensional cloud of trapped degenerate fermions. In Sec. VI we consider purely two-dimensional dipolar fermions to compare it with the quasi-two-dimensional case. In Sec. VII brief conclusions are presented.

\section{Self-consistent field model}

The many-particle quantum hydrodynamic method \cite{MaksimovTMP 1999}, \cite{MaksimovTMP 2001} for quantum gases with the short-range interaction was developed in Ref. \cite{Andreev PRA08}.
The method of derivation  of the quantum hydrodynamic equations for electric dipolar particles was presented in Ref. \cite{Andreev PRB 11}. It was applied to dipolar ultracold bosons in Refs. \cite{Andreev EPJ D 13}, \cite{Andreev MPL 13}. Similar derivation can be performed for ultracold dipolar fermions. Result of derivation for ultracold dipolar fermions with the evolution of dipole directions was presented in Ref. \cite{Andreev RPJ 13}. Applying the self-consistent field approximation we find the continuity and the Euler equations
\begin{equation}\label{Bscf_Fer cont eq}
\partial_{t}n+\nabla\cdot(n\textbf{v})=0 ,\end{equation}
and
$$mn(\partial_{t}+\textbf{v}\cdot\nabla)\textbf{v} +\nabla p -\frac{\hbar^{2}}{4m}n\nabla\Biggl(\frac{\triangle n}{n}-\frac{(\nabla n)^{2}}{2n^{2}}\Biggr)$$
\begin{equation}\label{Bscf_Fer bal imp eq} =P^{\beta}\nabla E_{ext}^{\beta}+P^{\beta}\nabla\int d\textbf{r}'G^{\beta\gamma}(\textbf{r},\textbf{r}')P^{\gamma}(\textbf{r}',t).\end{equation}
In equations (\ref{Bscf_Fer cont eq}) and (\ref{Bscf_Fer bal imp eq}) we have used the following notations: $n$ is the concentration of particles, $\textbf{v}$ is the velocity field, presenting dynamics of local center of mass, $\partial_{t}$ is the time derivative, $\nabla$ ($\triangle$) is the gradient (Laplace) operator, $m$ is the mass of particles, $\hbar$ is the Planck constant, $\textbf{E}_{ext}$ is the external electric field, $\textbf{P}$ is the polarisation, or, in other words, the density of the electric dipole moment, $p$ is the pressure (or the Fermi pressure), which is related to the distribution of particles over the quantum states, it depends on details of quantum state occupation, $G^{\alpha\beta}=\partial^{\alpha}\partial^{\beta}\frac{1}{|\textbf{r}-\textbf{r}'|}$ is the electric dipole-dipole interaction, which is a symmetric second rank tensor, and $d\textbf{r}'=dx dy dz$.

To get a closed set of the hydrodynamic we need to consider polarisation $\textbf{P}$ and the pressure $p$ (equation of state).

In the systems of the spin unpolarised degenerate spin-1/2 fermions, the pressure equals to the Fermi pressure $p=p_{Fe}$. The Fermi pressure for the three dimensional (3D) systems of particles is $p_{Fe,3D}=(3\pi^{2})^{2/3}\hbar^{2}n_{3D}^{5/3}/(5m)$, which contains the three dimensional  particle concentration $n=n_{3D}$, $[n_{3D}]=cm^{-3}$. In the case of the two-dimensional (2D) plane-like structures in the three dimensional space we have $p=p_{Fe,2D}=\pi\hbar^{2}n_{2D}^{2}/(2m)$ containing the two dimensional particle concentration $[n_{2D}]=cm^{-2}$.

For the fully spin polarized systems of electric dipoles
equations of state appears as
$p_{3D\uparrow\uparrow}=(6\pi^{2})^{2/3}\hbar^{2}n_{3D}^{5/3}/(5m)$
for 3D mediums, and
$p_{2D\uparrow\uparrow}=\pi\hbar^{2}n_{2D}^{2}/m$ for
2D mediums, where subindex $\uparrow\uparrow$ means that all
particles have same spin direction.

Next let us consider the equation state for the partially spin polarised spin-1/2 ultracold fermions possessing the large electric dipole moment, where we find $p=p_{3D\Updownarrow}=\vartheta_{3D}(3\pi^{2})^{2/3}\hbar^{2}n_{3D}^{5/3}/(5m)$
for 3D mediums, with
\begin{equation}\label{Bscf_Fer} \vartheta_{3D}=\frac{1}{2}[(1+\eta)^{5/3}+(1-\eta)^{5/3}],\end{equation}
where $\Updownarrow$ stands for partially polarized systems, that
means that part of states contain two particle with opposite spins
and other occupied states contain one particle with same spin
direction. Here we need to introduce ratio of spin
polarisability for system of spin-1/2 electric dipoles $\eta=\frac{\mid
n_{\uparrow}-n_{\downarrow}\mid}{n_{\uparrow}+n_{\downarrow}}$,
with indexes $\uparrow$ and $\downarrow$ means particles with spin
up and spin down.

Let us note that the pressure of the fully polarised spins larger than the pressure of the unpolarised fermions: $\frac{p_{\uparrow\uparrow,3D}}{p_{\uparrow\downarrow,3D}}=\sqrt[3]{4}$, and $\frac{p_{\uparrow\uparrow,2D}}{p_{\uparrow\downarrow,2D}}=2$.

The kinetic theory shows that the small perturbations of pressure, in the linear regime, differ from the results of application of the linearisation of the Fermi pressure by the factor $\chi_{K}=9/5$, for three dimensional mediums. To include this information we substitute coefficient $\chi_{K}$ in the pressure perturbations $p=p_{0}+\delta p$, and
\begin{equation}\label{Bscf_Fer pressure perturb from Kin Th}\delta p=\chi_{K}\delta p_{3D\Updownarrow}.\end{equation}

Polarisation $\textbf{P}$ simplifies in the case of the align dipoles.
In this case evolution of polarisation is reduced to the concentration evolution $\textbf{P}=nd\textbf{l}$, where $\textbf{l}$ is the unit vector in direction of the polarization formed by the external electric field. Approximations of the self-consistent field and the aligned dipoles allow to obtain a closed set of the QHD equations for dipolar ultracold fermions.

The self-consistent field approximation applied in equation (2) allows to introduce the intrinsic electric field created by dipoles
\begin{equation}\label{Bscf_Fer E explicit Integral} E^{\alpha}_{int}=d^{\beta}\int d\textbf{r}' G^{\alpha\beta}(\textbf{r},\textbf{r}')n(\textbf{r}',t).\end{equation}
Using notion of full electric field  $\textbf{E}_{full}=\textbf{E}_{ext}+\textbf{E}_{int}$ we can represent the Euler equation (2) in more simple non-integral form
$$mn(\partial_{t}+\textbf{v}\cdot\nabla)\textbf{v}+\nabla p$$
\begin{equation}\label{Bscf_Fer bal imp eq with E_full} -\frac{\hbar^{2}}{4m}n\nabla\Biggl(\frac{\triangle n}{n}-\frac{(\nabla n)^{2}}{2n^{2}}\Biggr) =P^{\beta}\nabla E_{full}^{\beta}.\end{equation}

The non-integral form of the Euler equation contains an extra variable: the electric field $\textbf{E}=\textbf{E}_{full}$. To get closed set of equations we need to obtain equations for electric field $\textbf{E}$.

Acting by the operators $\textrm{div}$ and $\textrm{curl}$ on the explicit form of electric field (\ref{Bscf_Fer E explicit Integral}) we show that the electric field satisfies the Maxwell equations
\begin{equation}\label{Bscf_Fer field good div}\nabla\cdot\textbf{E}(\textbf{r},t)=-4\pi \nabla\cdot\textbf{P}(\textbf{r},t)=-4\pi d(\textbf{l}\cdot\nabla) n(\textbf{r},t),\end{equation}
and
\begin{equation}\label{Bscf_Fer field good curl}\nabla\times\textbf{E}(\textbf{r},t)=0.\end{equation}

In the limit of the aligned dipoles the force field of the dipole-dipole interaction and the interaction of the dipoles with the external electric field can be rewritten as $nd\partial^{\alpha}(\textbf{l}\textbf{E})$,
where $\textbf{l}$ is the unit vector in direction of the polarization formed by the external electric field.

Under assumption of the potential velocity field $\textbf{v}=\nabla\phi$, the set of quantum hydrodynamic equations (\ref{Bscf_Fer cont eq}) and (\ref{Bscf_Fer bal imp eq with E_full}) can be presented in the form of the non-linear Schrodinger equation for the effective many-particle wave function $\Phi=\sqrt{n}\exp(\imath m\phi/\hbar)$:
\begin{equation}\label{Bscf_Fer nlse int polariz some frame non Int}\imath\hbar\partial_{t}\Phi(\textbf{r},t)=\Biggl(-\frac{\hbar^{2}}{2m}\triangle +\vartheta_{3D}\frac{(3\pi^{2})^{2/3}\hbar^{2}n^{2/3}}{2m}-\textbf{d}\cdot\textbf{E}\Biggr)\Phi(\textbf{r},t),\end{equation}
$n(\textbf{r},t)=\Phi^{*}(\textbf{r},t)\Phi(\textbf{r},t)$ is the concentration of particles, and $\textbf{d}=d\textbf{l}$. The first term on the right-hand side of the NLSE (\ref{Bscf_Fer nlse int polariz some frame non Int}) is the kinetic energy operator. The second term presents the Fermi pressure of the partially spin polarised systems of spin-1/2 fermions possessing the electric dipole moment. The last term describes the potential energy of the electric dipoles being in the electric field $\textbf{E}$ consisting of the external electric field and the internal electric field created by the dipoles. Formula (\ref{Bscf_Fer nlse int polariz some frame non Int}) gives the NLSE for dipolar fermions in the self-consistent field approximation.

Definitions of the particle concentration $n$, the velocity field $\textbf{v}$, and the polarisation $\textbf{P}$ are
\begin{equation}\label{Bscf_Fer def density}n(\textbf{r},t)=\int \textrm{d}\textrm{R}_{N}\sum_{i}\delta(\textbf{r}-\textbf{r}_{i})\psi^{*}(R,t)\psi(R,t),\end{equation}
\begin{equation}\label{Bscf_Fer polarisation Def} \textbf{P}(\textbf{r},t)=\int \textrm{d}\textrm{R}_{N}\sum_{i}\delta(\textbf{r}-\textbf{r}_{i})\textbf{d}_{i}\psi^{*}(R,t)\psi(R,t), \end{equation}
and
$$\textbf{v}(\textbf{r},t)=\frac{\textbf{j}(\textbf{r},t)}{n(\textbf{r},t)} =\int \textrm{d}\textrm{R}_{N}\sum_{i}\delta(\textbf{r}-\textbf{r}_{i})\times$$
\begin{equation}\label{Bscf_Fer def current} \times\biggl(\psi^{*}(R,t)\textbf{p}_{i}\psi(R,t)+c.c.\biggr),\end{equation}
correspondingly.

Definitions (\ref{Bscf_Fer def density})-(\ref{Bscf_Fer def current}) are presented in terms of the microscopic many-particle wave function $\psi(R,t)$. The wave function $\psi(R,t)$ obeys the many-particle Schrodinger equation. We do not consider the spin evolution assuming that all atoms or molecules in the same fine structure state, hence we do not need to use the Pauli equation. Consequently Schrodinger equation, corresponding to the systems under consideration, within the
quasi-static approximation, has the following form
\begin{equation}\label{Bscf_Fer Schrodinger} \imath\hbar\partial_{t}\psi=\hat{H}\psi,\end{equation}
with
$$\hat{H}=\sum_{i}\Biggl(\frac{1}{2m_{i}}\hat{\textbf{p}}_{i}^{2}-\textbf{d}_{i}\textbf{E}_{i,ext}+V_{trap}(\textbf{r}_{i},t)\Biggr)$$
\begin{equation}\label{Bscf_Fer Hamiltonian}+\frac{1}{2}\sum_{i,j\neq i}\Biggl(U_{ij}-d_{i}^{\alpha}d_{j}^{\beta}G_{ij}^{\alpha\beta}\Biggr).\end{equation}
The first term in the Hamiltonian is the operator of the kinetic energy. The second
term represents the interaction between the dipole moment
$d_{i}^{\alpha}$ and the external electrical field. The subsequent
terms represent the short-range $U_{ij}$ and the dipole-dipole $d_{i}^{\alpha}d_{j}^{\beta}G_{ij}^{\alpha\beta}$
interactions between particles. The Green function
for the dipole-dipole interaction reads as
$G_{ij}^{\alpha\beta}=\nabla^{\alpha}_{i}\nabla^{\beta}_{i}(1/r_{ij})$. This Schrodinger equation coincides with the Schrodinger equation applied for dipolar bosons dynamics in Refs. \cite{Andreev EPJ D 13}, \cite{Andreev RPJ 12}, \cite{Andreev JP B 14}, \cite{Andreev EPJ D 14}.

Equation (\ref{Bscf_Fer Schrodinger}) shows that we do not apply the second quantization. The wave function $\psi(R,t)$ governs the microscopic evolution of the systems of interacting particles.

The explicit form of the Green function of the dipole-dipole interaction can be written in different forms:
$$\textrm{G}^{\alpha\beta}(\textbf{r},\textbf{r}')=\partial^{\alpha}\partial^{\beta}\frac{1}{|\textbf{r}-\textbf{r}'|}$$
$$ =-\frac{\delta^{\alpha\beta}-3r^{\alpha}r^{\beta}/r^{2}}{r^{3}}-\frac{4\pi}{3}\delta^{\alpha\beta}\delta(\textbf{r})$$
\begin{equation}\label{Bscf_Fer dd int Green funct} =-\frac{\delta^{\alpha\beta}-3r^{\alpha}r^{\beta}/r^{2}}{r^{3}}+\frac{1}{3}\delta^{\alpha\beta}\triangle\frac{1}{r}.\end{equation}
Explicit form of the Green function of the electric dipole-dipole interaction consists of two parts: the reduced part (the first term in the second line of formula (\ref{Bscf_Fer dd int Green funct})) and the delta function term. The full theory, which is in accordance with the Maxwell equations, requires the application of the full potential of the dipole-dipole interaction containing the delta-function term (\ref{Bscf_Fer dd int Green funct}).

Below, at the consideration of the two dimensional dipolar fermions we will need to apply the explicit form of the $zz$ element of the tensor of Green function of the electric dipole interaction $G^{zz}$. Hence we find this element from the formula (\ref{Bscf_Fer dd int Green funct}):
$$G^{zz}(\xi)=\textrm{G}^{\alpha\beta}(\xi)\delta^{z\alpha}\delta^{z\beta}$$
$$=\biggl(-\frac{\delta^{\alpha\beta}}{\xi^{3}} +\frac{3\xi^{\alpha}\xi^{\beta}}{\xi^{5}} -\frac{4\pi}{3}\delta^{\alpha\beta}\delta(\xi)\biggr)\delta^{z\alpha}\delta^{z\beta}$$
\begin{equation}\label{Bscf_Fer dd int Green funct zz} =-\frac{1}{\xi^{3}}+\frac{3(\xi^{z})^{2}}{\xi^{5}} -\frac{4\pi}{3}\delta(\xi).\end{equation}

\section{Dipolar fermions beyond the self-consistent field approximation: Exchange dipole-dipole interaction}

General form of the Euler equation beyond the self-consistent field approximation, which arises at derivation \cite{Andreev PRB 11}, is
$$mn(\partial_{t}+\textbf{v}\cdot\nabla)\textbf{v}+\nabla p-\frac{\hbar^{2}}{4m}n\nabla\Biggl(\frac{\triangle n}{n}-\frac{(\nabla n)^{2}}{2n^{2}}\Biggr)$$
\begin{equation}\label{Bscf_Fer bal imp eq with P2}= -\Sigma+
\textrm{P}^{\beta}\nabla \textrm{E}^{\beta}_{ext}+\int d\textbf{r}'(\nabla \textrm{G}^{\beta\gamma}(|\textbf{r}-\textbf{r}'|)) \textrm{P}_{2}^{\beta\gamma}(\textbf{r},\textbf{r}',t),\end{equation}
where the first term on the right-hand side presents the general form of the short-range interaction $\Sigma^{\alpha}=\partial^{\beta}\sigma^{\alpha\beta}$, with the quantum stress tensor $\sigma^{\alpha\beta}$ \cite{Andreev PRA08}, \cite{Andreev IJMP B 13}, the last term on the right-hand side describes the dipole-dipole interaction. The quantum stress tensor appears as an expansion in the series on the interaction radius of the short-range interaction $\sigma^{\alpha\beta}$ \cite{Andreev PRA08}. In systems of bosons, being in the Bose-Einstein condensate state, a contribution of the short range interaction arises in the first order by the interaction radius, if we consider spherically symmetric short-range potential. It leads to the Gross-Pitaevskii approximation \cite{Andreev PRA08}. Generalization of the Gross-Pitaevskii model in the third order by the interaction radius can be found in Ref. \cite{Andreev PRA08}. Influence of the short interaction considered up to the third order by the interaction radius on spectrum of collective excitations in dipolar Bose-Einstein condensates and dipolar fermions was considered in Refs. \cite{Andreev EPJ D 13} and \cite{Andreev RPJ 13} correspondingly. If we consider fermions then the contribution of the short-range interaction equals to zero due to the antisymmetry of the wave function of fermions. However, a non-zero contribution was derived in the third order by the interaction radius \cite{Andreev PRA08}. Including the dependence on the spin polarisation, the result of Ref. \cite{Andreev PRA08} can be presented as $\sigma^{\alpha\beta}=\frac{ 4m}{\hbar^{2}}\Upsilon_{2}[\delta^{\alpha\beta}\vartheta_{3D}(3\pi^{2})^{\frac{2}{3}}\frac{1}{m}n^{\frac{8}{3}}+\delta^{\alpha\beta}n\tilde{T}^{\gamma\gamma}
+2n\tilde{T}^{\alpha\beta}]$, where $\tilde{T}^{\alpha\beta}=-\frac{1}{4m}[\partial^{\alpha}\partial^{\beta}n-\frac{1}{n}(\partial^{\alpha}n)(\partial^{\beta}n)]$, and $\Upsilon_{2ij}=\frac{4\pi}{15} \int{dr r^5 \frac{\partial U_{ij}
(r)}{\partial r}}$.

This is the explicit definition of the two-particle concentration in terms of the many-particle wave function occurring at derivation of the Euler equation
$$\textrm{P}_{2}^{\alpha\beta}(\textbf{r},\textbf{r}',t)=\int \textrm{d}\textrm{R}_{N}\sum_{i,j\neq i}\delta(\textbf{r}-\textbf{r}_{i})\delta(\textbf{r}'-\textbf{r}_{j})\times$$
\begin{equation}\label{Bscf_Fer polarisation 2part Def}\times d_{i}^{\alpha}d_{j}^{\beta}\psi^{*}(R,t)\psi(R,t). \end{equation}

In this paper we focus our attention on dipolar fermions with aligned dipoles. Consequently polarisation of the system reduces to the concentration
\begin{equation}\label{Bscf_Fer}\textbf{P}=d\textbf{l}n, \end{equation}
with $\textbf{l}$ direction of all dipoles parallel to the external electric field. Let us choose $\textbf{l}=\textbf{e}_{z}$.

For the aligned dipoles the two-particle polarisation can be simplified as
\begin{equation}\label{Bscf_Fer} \textrm{P}_{2}^{\alpha\beta}(\textbf{r},\textbf{r}',t)=d^{2}\delta^{\alpha z}\delta^{\beta z} n_{2}(\textbf{r},\textbf{r}',t), \end{equation}
where the two-particle concentration appears to be
$$n_{2}(\textbf{r},\textbf{r}',t)$$
\begin{equation}\label{Bscf_Fer concentration 2part Def} =\int \textrm{d}\textrm{R}_{N}\sum_{i,j\neq i}\delta(\textbf{r}-\textbf{r}_{i})\delta(\textbf{r}'-\textbf{r}_{j})\psi^{*}(R,t)\psi(R,t). \end{equation}

General form of the two-particle concentration for fermions was obtained in Refs. \cite{MaksimovTMP 1999}, \cite{Andreev PRA08}
\begin{equation}\label{Bscf_Fer n2 expansion} n_2(\textbf{r},\textbf{r}',t)=n(\textbf{r},t)n(\textbf{r}',t)
-|\rho(\textbf{r},\textbf{r}',t)|^{2},\end{equation}
where,
\begin{equation}\label{Bscf_Fer n   via varphi}
n(\textbf{r},t)=\sum_{g}n_{g}\varphi_{g}^{*}(\textbf{r},t)\varphi_{g}(\textbf{r},t)
,\end{equation}
and
\begin{equation}\label{Bscf_Fer rho via varphi} \rho(\textbf{r},\textbf{r}',t)=\sum_{g}n_{g}\varphi_{g}^{*}(\textbf{r},t)\varphi_{g}(\textbf{r}',t),\end{equation}
with $\varphi_{g}(\textbf{r},t)$ are the
arbitrary single-particle wave functions.

Exchange electric dipole-dipole interaction in system of spin-1/2 fully spin polarised degenerate fermions of align electric dipoles is
\begin{equation}\label{Bscf_Fer force of exchange in fully polarised} F_{SHF,3D,Exc}=\frac{8\pi}{3}d^{2} n\nabla n.\end{equation}
Formula (\ref{Bscf_Fer force of exchange in fully polarised}) introduce an attractive interaction.

Including of the partial spin polarisation degenerate fermions of align electric dipoles gives
\begin{equation}\label{Bscf_Fer} F_{SHF,3D,Exc}=\eta\frac{8\pi}{3}d^{2} n\nabla n,\end{equation}
where
\begin{equation}\label{Bscf_Fer} \eta=\frac{\mid
n_{\uparrow}-n_{\downarrow}\mid}{n_{\uparrow}+n_{\downarrow}}.\end{equation}

Let us present the explicit form of the Euler equation for 3D degenerate fully polarised dipolar fermions with the exchange dipole-dipole interaction
$$mn(\partial_{t}+\textbf{v}\cdot\nabla)\textbf{v}+\sqrt[3]{2}\nabla p_{Fe}-\frac{\hbar^{2}}{4m}n\nabla\Biggl(\frac{\triangle n}{n}-\frac{(\nabla n)^{2}}{2n^{2}}\Biggr)$$
\begin{equation}\label{Bscf_Fer bal imp eq SCF with exchange}=
\textrm{P}^{\beta}\nabla \textrm{E}^{\beta}_{full}+\frac{8\pi}{3}d^{2} n\nabla n.\end{equation}

Corresponding non-linear Schrodinger equation arises as
$$\imath\hbar\partial_{t}\Phi(\textbf{r},t)=\Biggl(-\frac{\hbar^{2}}{2m}\triangle+\sqrt[3]{4} \frac{(3\pi^{2})^{2/3}\hbar^{2}n^{2/3}}{2m}$$
\begin{equation}\label{Bscf_Fer nlse with exchange} -\textbf{d}\cdot\textbf{E}-\frac{8\pi}{3}d^{2} n\Biggr)\Phi(\textbf{r},t).\end{equation}
Formula (\ref{Bscf_Fer nlse with exchange}) presents the NLSE for the fully spin polarised fermions with the aligned electric dipoles with the account of the exchange dipole-dipole interaction. The exchange interaction is presented by the last term on the right-hand side of the NLSE.

\begin{figure}
\includegraphics[width=8cm,angle=0]{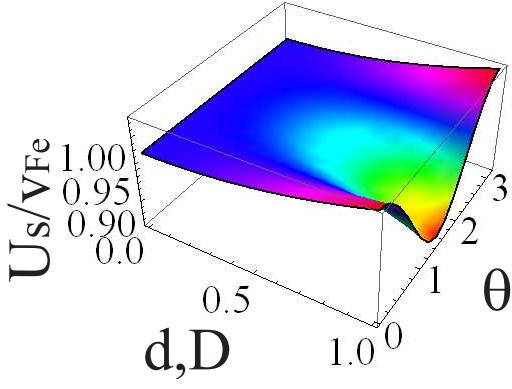}
\caption{\label{BF01} (Color online) The figure shows the dimensionless sound velocity $U=U_{s,3D}/v_{Fe,3D}$ as the function of the electric dipole moment $d$, in Debay (D) units, and the angle $\theta$ between the direction of the external field and the direction of wave propagation at $n_{0}=10^{15}cm^{-3}$ and mass of particles $m=127$ amu (the atomic mass unit).}
\end{figure}
\begin{figure}
\includegraphics[width=8cm,angle=0]{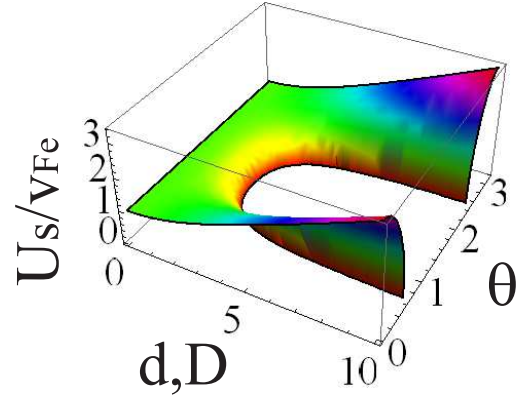}
\caption{\label{BF02} (Color online) The figure shows the appearance of instability for the dipolar fermions possessing the electric dipole moment more that 2 Debay at $n_{0}=10^{15}cm^{-3}$ and mass of particles $m=127$ amu. Area of instability faster increases at $d\in(2,4)$ Debay reaching angle $\theta_{0}\approx0.61$ radian.}
\end{figure}
\begin{figure}
\includegraphics[width=8cm,angle=0]{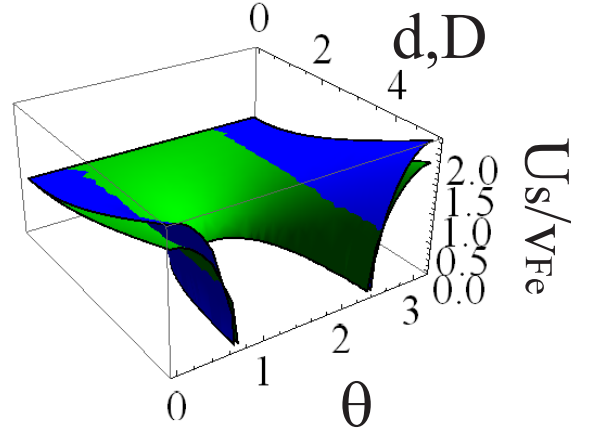}
\caption{\label{BF03} (Color online) The figure shows behavior of the sound velocity $U_{s,3D}(d,\theta)$ for particles of different mass ($m=127$ amu (the green surface) and $m=200$ amu (the blue surface)) at $n_{0}=10^{15}cm^{-3}$.}
\end{figure}

\section{Collective excitations in 3D unlimited sample}

Our calculation gives following spectrum of collective excitations
$$\omega^{2}=\frac{4\pi n_{0}d^{2}k^{2}}{m}\cos^{2}\theta-\eta\frac{8\pi n_{0}d^{2}k^{2}}{3m}$$
\begin{equation}\label{Bscf_Fer dispersion 3D} +\vartheta_{3D}\chi_{K}\frac{(3\pi^{2})^{2/3}\hbar^{2}n_{0}^{2/3}}{3m^{2}}k^{2}+\frac{\hbar^{2}k^{4}}{4m^{2}},\end{equation}
where $\cos\theta=k_{z}/k$.

At small wave vectors $k$ and the full spin polarisation the frequency can be written as
\begin{equation}\label{Bscf_Fer dispersion 3D LongWaveL}\omega =\sqrt{\chi_{K}\frac{(6\pi^{2})^{2/3}\hbar^{2}n_{0}^{2/3}}{3m^{2}} +\frac{4\pi n_{0}d^{2}}{m}\biggl(\cos^{2}\theta-\frac{2}{3}\biggr)}k. \end{equation}

Without the dipole-dipole interaction formula (\ref{Bscf_Fer dispersion 3D LongWaveL}) gives $\omega=U_{s,3D}k$, where $U_{s,3D}$ is the sound velocity of ideal Fermi gas caused by the Fermi pressure: $U_{s,3D}=\sqrt{\chi_{K}}\frac{\sqrt[3]{2}}{\sqrt{3}}v_{Fe,3D}$, with the traditional Fermi velocity $v_{Fe,3D}=(3\pi^{2}n_{0,3D})^{1/3}\hbar/m$.

For numerical analysis of the spectrum (\ref{Bscf_Fer dispersion 3D LongWaveL}) we apply a fixed value of the equilibrium concentration $n_{0}=10^{15}cm^{-3}$ and mass of particles $m=127$ amu.
Fig. (\ref{BF01}) shows spectrum (\ref{Bscf_Fer dispersion 3D LongWaveL}), or the rate of the sound velocity to the Fermi velocity, at the small electric dipole moments $d<1$ Debay (D). Fig. (\ref{BF02}) shows the arising of the instability at the larger electric dipole moments. The area of instability is shown by the clipping of the surface. Change of spectrum at the increase of the mass of particles from $m=127$ amu (the green surface) to $m=200$ amu (the blue surface) is presented on Fig. (\ref{BF03}).

\section{Quasi two-dimensional dipolar Fermi gas with exchange interaction}

\subsection{Transformation of non-local terms}

We have considered homogeneous three dimensional systems of dipolar degenerate fermions. In this section we consider fermions being in a trap, we choose to consider quasi-two dimensional trap with the strong confinement along z-direction.

In this section we follow Refs. \cite{Fischer PRA 06R}, \cite{Petrov PRL 00}, where method of consideration of quasi two-dimensional clouds of ultracold gases. Quasi two-dimensional dipolar BECs were considered in Ref. \cite{Fischer PRA 06R}.

Before we perform transformation of the NLSE in the form corresponding to the quasi-two-dimensional trap, let us represent the NLSE in the suitable form. Here we apply the NLSE for fully spin-polarised fermions possessing the electric dipole moment, which includes the exchange dipole-dipole interaction (\ref{Bscf_Fer nlse with exchange}). First we rewrite it in an integral form
$$\imath\hbar\partial_{t}\Phi(\textbf{r},t)=\Biggl(-\frac{\hbar^{2}}{2m}\triangle+\frac{(6\pi^{2})^{2/3}\hbar^{2}n^{2/3}}{2m}$$
\begin{equation}\label{Bscf_Fer nlse with exchange Q2D st1} -d^{\beta}d^{\gamma}\int d\textbf{r}'G^{\beta\gamma}(\textbf{r},\textbf{r}')n(\textbf{r}',t)-\frac{8\pi}{3}d^{2} n\Biggr)\Phi(\textbf{r},t).\end{equation}
Next we use the explicit form of the Green function of the electric dipole interaction (\ref{Bscf_Fer dd int Green funct})
$$\imath\hbar\partial_{t}\Phi(\textbf{r},t)=\Biggl(-\frac{\hbar^{2}}{2m}\triangle+\frac{(6\pi^{2})^{2/3}\hbar^{2}n^{2/3}}{2m}$$
$$+d^{\beta}d^{\gamma}\int d\textbf{r}'\frac{\delta^{\alpha\beta}-3\Delta r^{\alpha} \Delta r^{\beta}/(\Delta r)^{2}}{(\Delta r)^{3}}n(\textbf{r}',t)$$
\begin{equation}\label{Bscf_Fer nlse with exchange Q2D st2} +\frac{4\pi}{3}d^{2} n-\frac{8\pi}{3}d^{2} n\Biggr)\Phi(\textbf{r},t),\end{equation}
where $\Delta \textbf{r}=\textbf{r}-\textbf{r}'$, $\Delta r=\mid\Delta \textbf{r}\mid$, and we have taken integral with the term proportional to the Dirac delta function. So, two last terms have similar form. Combining the two last terms together and assuming that all electric dipoles are perpendicular to the plane of confinement (parallel to the z-direction) we find
$$\imath\hbar\partial_{t}\Phi(\textbf{r},t)=\Biggl(-\frac{\hbar^{2}}{2m}\triangle+\frac{(6\pi^{2})^{2/3}\hbar^{2}n^{2/3}}{2m}$$
\begin{equation}\label{Bscf_Fer nlse with exchange Q2D st3} +d^{2}\int d\textbf{r}'\frac{1-3\cos^{2}\vartheta_{\textbf{r}}}{(\Delta r)^{3}}n(\textbf{r}',t)-\frac{4\pi}{3}d^{2} n\Biggr)\Phi(\textbf{r},t),\end{equation}
where $\vartheta_{\textbf{r}}$ is the angle between $\Delta \textbf{r}$ and the z-direction. The last term in equation (\ref{Bscf_Fer nlse with exchange Q2D st3}) can be represented via effective interaction constant $g_{eff}=-4\pi d^{2}/3$.

Now we are going to find the NLSE for quasi-two-dimensional electric dipolar fermions. We ready to apply results of Ref. \cite{Fischer PRA 06R} to reach our goal.

\begin{equation}\label{Bscf_Fer concentration r with trap on z} n(\textbf{r})=\mid\Phi(\textbf{r})\mid^{2}=\frac{1}{\sqrt{\pi a_{z}^{2}}}\exp\biggl(-\frac{z^{2}}{a_{z}^{2}}\biggr)n(x,y).\end{equation}

In accordance with equation (\ref{Bscf_Fer nlse with exchange Q2D st3}) and Ref. \cite{Fischer PRA 06R} we need to consider the reduced potential of dipole-dipole interaction
\begin{equation}\label{Bscf_Fer dd pot energy Reduced} \tilde{V}_{dd}(\textbf{r})=d^{2}\frac{1-3z^{2}/r^{2}}{r^{3}}.\end{equation}
The Fourier transform of the dipole-dipole interaction (\ref{Bscf_Fer dd pot energy Reduced}) takes the form
\begin{equation}\label{Bscf_Fer} \tilde{V}_{dd}(\textbf{k})=\frac{4\pi}{3}d^{2}\biggl(\frac{3k_{z}^{2}}{k^{2}}-1\biggr),\end{equation}
with $k^{2}=k_{x}^{2}+k_{y}^{2}+k_{z}^{2}$.

The Fourier transform of the concentration (\ref{Bscf_Fer concentration r with trap on z}) appears as
\begin{equation}\label{Bscf_Fer} n(\textbf{k})=\exp\biggl(-\frac{1}{4}k_{z}^{2}a_{z}^{2}\biggr)n(k_{x},k_{y}).\end{equation}

The Fourier transform is calculated in accordance with the formulae
\begin{equation}\label{Bscf_Fer} n(\textbf{r})=\frac{1}{(2\pi)^{3}}\int d\textbf{k}e^{-\imath \textbf{k}\textbf{r}}n(\textbf{k}),\end{equation}
and
\begin{equation}\label{Bscf_Fer} n(\textbf{k})=\int d\textbf{r}e^{\imath \textbf{k}\textbf{r}}n(\textbf{r}).\end{equation}

The Hamiltonian of the reduced potential of the dipole-dipole interaction is
$$H_{dd(R)}=\frac{1}{2}\int d\textbf{r}\int d\textbf{r}' n(\textbf{r})\tilde{V}_{dd}(\textbf{r}-\textbf{r}')n(\textbf{r}')$$
$$=\frac{1}{2}\frac{1}{(2\pi)^{3}}\int d\textbf{k} n(\textbf{k})\tilde{V}_{dd}(\textbf{k})n(-\textbf{k})$$
$$=\frac{2\pi d^{2}}{3}\frac{1}{(2\pi)^{2}}\int dk_{x}dk_{y}n(k_{x},k_{y})n(-k_{x},-k_{y})\times$$
\begin{equation}\label{Bscf_Fer} \times\biggl[\frac{2}{\sqrt{2\pi}a_{z}}-\frac{3}{2}\exp\biggl(\frac{k_{2D}^{2}a_{z}^{2}}{2}\biggr)k_{2D}\textrm{erfc}\biggl(\frac{k_{2D}a_{z}}{\sqrt{2}}\biggr)\biggr],\end{equation}
where $\textrm{erfc}(z)=1-\textrm{erf}(z)$$=1-(2/\sqrt{\pi})\int_{0}^{z}\exp(-t^{2})dt$, $k_{2D}=\sqrt{k_{x}^{2}+k_{y}^{2}}$.

\begin{figure}
\includegraphics[width=8cm,angle=0]{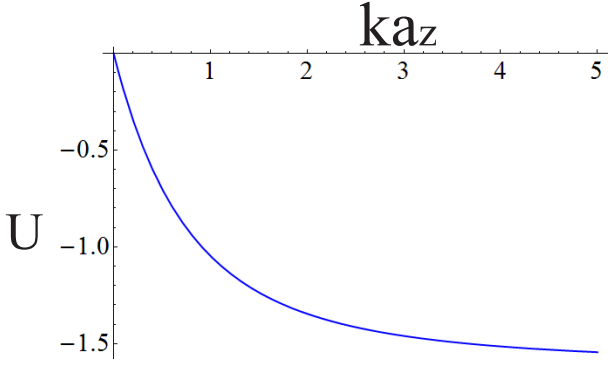}
\caption{\label{BF04} (Color online) The figure shows the dimensionless quasi two-dimensional Fourier image of the potential of electric dipole-dipole interaction of fermions $U=a_{z}\tilde{V}^{2D}_{tot}(\xi)/(\pi d^{2})$ as a function of the dimensionless module of two dimensional wave vector $ka_{z}$, where we include both the self-consistent field part and the exchange part of the dipole-dipole interaction.}
\end{figure}

Quasi two-dimensional Fourier image of all parts of the dipole-dipole interaction, the full self-consistent field part of the electric dipole-dipole interaction and the exchange part of the dipole-dipole interaction, which were represented as the sum of the reduced dipole-dipole potential and the effective short-range interaction, appears as follows
$$\tilde{V}^{2D}_{tot}(\xi)=\frac{g_{eff}}{2a_{z}}+\frac{2\pi d^{2}}{3a_{z}}[1-3\xi w(\xi/\sqrt{2})]$$
\begin{equation}\label{Bscf_Fer Fourier Tr of full DD Int} =\frac{\pi d^{2}}{a_{z}}[-2\xi w(\xi/\sqrt{2})], \end{equation}
with $\xi=k_{2D}a_{z}$, $k_{2D}=\sqrt{k_{x}^{2}+k_{y}^{2}}$, and $w(x)=\exp(x^{2})\textrm{erfc}(x)$. The dimensionless quasi two-dimensional Fourier image of  the dipole-dipole interaction $U=a_{z}\tilde{V}^{2D}_{tot}(\xi)/(\pi d^{2})$ is shown on Fig. (\ref{BF04}).

If we want to obtain corresponding NLSE we need to perform reverse two-dimensional Fourier transform of $\tilde{V}^{2D}_{tot}(\xi)$. However we can use result (\ref{Bscf_Fer Fourier Tr of full DD Int}) to find the contribution of the electric dipole-dipole interaction in the spectrum of linear collective excitations in quasi-two-dimensional dipolar fermions.

\subsection{Transformation of local terms at transition to quasi two-dimensional regime}

In our model the combination of the part of the self-consistent field dipole-dipole interaction containing the delta function and the exchange dipole-dipole interaction give a term in the NLSE, which can be considered as a local interaction. Moreover this term has structure similar to the short-range interaction of bosons in the Gross-Pitaevskii model. In the case of local interaction we can use more simple procedure to perform the transition of the Hamiltonian to the quasi two-dimensional regime. We can simply integrate over the confined direction
$$H_{SRI}=\frac{1}{2}\int d\textbf{r} g_{3D}n^{2}(\textbf{r})$$
$$=\frac{1}{2}g_{3D}\frac{1}{\pi a_{z}^{2}}\int d\textbf{r}_{2d} dz e^{-2\frac{z^{2}}{a_{z}^{2}}}n^{2}(x,y)$$
\begin{equation}\label{Bscf_Fer} =\frac{1}{2}g_{2D}\int d\textbf{r}_{2d} n^{2}(x,y),\end{equation}
where $d\textbf{r}_{2d}=dxdy$, and $g_{2D}=\frac{g_{3D}}{\sqrt{2\pi}a_{z}}$.

Corresponding NLSE arises as
\begin{equation}\label{Bscf_Fer} \imath\hbar\partial_{t}\Phi(x,y,t)=\Biggl(-\frac{\hbar^{2}}{2m}\triangle_{2D}+g_{2D} n\Biggr)\Phi(x,y,t).\end{equation}

Similarly we can consider the Fermi pressure. Corresponding Hamiltonian and following calculations are
$$H_{F}=\frac{3}{5}\int d\textbf{r} C_{3D}n^{5/3}(\textbf{r})$$
$$=\frac{3}{5}C_{3D}\int d\textbf{r}_{2D}dz n^{5/3}(x,y)\biggl(\frac{1}{\sqrt{\pi a_{z}^{2}}}e^{-\frac{z^{2}}{a_{z}^{2}}}\biggr)^{5/3}$$
\begin{equation}\label{Bscf_Fer} =\frac{3}{5}C_{2D}\int d\textbf{r}_{2D} n^{5/3}(x,y),\end{equation}
where $C_{3D}=\vartheta_{3D}(3\pi^{2})^{2/3}\hbar^{2}/(2m)$, and $C_{2D}=\sqrt{\frac{3}{5}}\frac{1}{(\pi a_{z}^{2})^{1/3}}C_{3D}=\vartheta_{3D}\sqrt{\frac{3}{5}}\frac{9^{1/3}\pi\hbar^{2}}{2ma_{z}^{2/3}}$.

Corresponding the NLSE, for fermions with the Fermi pressure in absence of the interaction, in quasi-two-dimensional case appears as
$$\imath\hbar\partial_{t}\Phi(x,y,t)=\Biggl(-\frac{\hbar^{2}}{2m}\triangle_{2D}$$
\begin{equation}\label{Bscf_Fer} +\vartheta_{3D} \sqrt{\frac{3}{5}}\frac{9^{1/3}\pi\hbar^{2}}{2ma_{z}^{2}}n^{2/3}\Biggr)\Phi(x,y,t)\end{equation}

Quasi-two-dimensional Fermi pressure appears as
\begin{equation}\label{Bscf_Fer} p_{Fe(q2D)}=\vartheta_{3D}\chi_{K}\sqrt{\frac{3}{5}}\frac{3^{2/3}\pi\hbar^{2}}{5ma_{z}^{2}}n_{q2D}^{5/3}.\end{equation}
Correspondingly, for the quasi-two-dimensional sound velocity of ideal fermi gas we have
\begin{equation}\label{Bscf_Fer} U_{s(q2D)}=\sqrt{\vartheta_{3D}\chi_{K}}\frac{\sqrt[3]{3}\sqrt{\pi}}{\sqrt[4]{15}}\frac{\hbar}{ma_{z}}n_{q2D,0}^{1/3}.\end{equation}

\subsection{Spectrum of collective excitations in quasi two-dimensional dipolar Fermi gas}

Dipolar part of the spectrum of the linear collective excitations is proportional to $\tilde{V}^{2D}_{tot}(k_{2D}a_{z})$.

The full spectrum also contains the contribution of the Fermi pressure and the quantum Bohm potential.

For small $\xi$, function $w(\xi/\sqrt{2})$ behaves like $w(\xi/\sqrt{2})=1-\sqrt{2/\pi}\xi+\emph{O}(\xi^{2})$.

Spectrum in the small wave vector regime, containing terms up to $k_{2D}^{4}$ can be presented as follows
$$\omega^{2}=\frac{\pi\sqrt{\pi}n(0)d^{2}}{m}k_{2D}^{2}\biggl[-2k_{2D}a_{z}+2\sqrt{\frac{2}{\pi}}k_{2D}^{2}a_{z}^{2}\biggr]$$
\begin{equation}\label{Bscf_Fer spectrum quasi2D small k} +U_{s(q2D)}^{2}k_{2D}^{2}+\frac{\hbar^{2}k_{2D}^{4}}{4m^{2}}.\end{equation}
The first group of terms in formula (\ref{Bscf_Fer spectrum quasi2D small k}) describes the contribution of the dipole-dipole interaction in the small wave vector limit. It consists of two parts. The first of them is negative and proportional to $k^{3}$. Another term is positive and proportional to $k^{4}$ as the quantum Bohm potential presented by the last term in formula (\ref{Bscf_Fer spectrum quasi2D small k}). The second term in formula (\ref{Bscf_Fer spectrum quasi2D small k}) presents the Fermi pressure, and it is proportional to $k^{2}$.

We see that the Fermi pressure leads to the sound like spectrum at the small $k$. Moreover, the account of the exchange part of the electric dipole-dipole interaction in the fully spin polarised spin-1/2 fermions reveals in the fact that the dipole-dipole interaction does not show sound like spectrum. The contribution of the dipole-dipole interaction is proportional to the third degree of the module of wave vector and to higher orders of the module of wave vector.

Parameter $R$ introduced in Ref. \cite{Fischer PRA 06R}, in our case, has the magnitude of
\begin{equation}\label{Bscf_Fer} R\equiv\frac{\sqrt{\pi/2}}{1+\frac{g_{3D}}{2g_{d}}}=2.51,\end{equation}
where $g_{3D}=g_{eff}=-4\pi d^{2}/3$, $g_{d}=4\pi d^{2}/3$.

\section{Two-dimensional dipolar Fermi gas}

In the previous section we have considered the quasi two-dimensional dipolar fermions, where we have explicitly considered confinement of particles by the harmonic trap. At consideration of the reduced potential of the dipole-dipole interaction we have followed Ref. \cite{Fischer PRA 06R}, where the quasi two-dimensional dipolar Bose-Einstein condensates were considered for the first time. Analysis of quasi two-dimensional dipolar fermions was performed in Ref. \cite{Lima PRA 10}.
Two dimensional dipolar BECs were recently considered in Refs. \cite{Lu_Shlyapnikov arX_14} and \cite{Boudjemaa PRA 13}, where authors apply the purely two-dimensional Fourier image for ultrathin plane of dipolar fermions with no trace of the delta function term in the potential of electric dipole interaction. Similarly to earlier works on three-dimensional dipolar Bose-Einstein condensates \cite{Santos PRL 00}, \cite{Goral PRA 00}, \cite{Yi PRA 00}. Purely two-dimensional models are widely spread in the condensed matter physics at description of the two-dimensional electron gas (2DEG) (see for instance \cite{Fetter AoP 73}-\cite{Batke PRB 86}).

In previous section we have considered three dimensional dipolar fermions in traps with the strong confinement in the z-direction and explicit account of the trapping in the z-direction. It have given us the quasi-two dimensional distribution of particles. However, there is an approach to consider ultrathin plane-like structures as purely two dimensional objects in the three dimensional space. Let us apply this approach to derive a model of the two-dimensional dipolar fermions with the exchange electric dipole-dipole interaction.

We can start our analysis with the Schrodinger equation in the plane-like two dimensional layer. As for three dimensional case we give definition of the two-dimensional concentration of particles
\begin{equation}\label{Bscf_Fer def density 2D} n_{2D}(\textbf{r},t)=\int dR_{2N} \sum_{i}\delta(\textbf{r}_{2D}-\textbf{r}_{2D,i})\psi^{*}(R_{2N},t)\psi(R_{2N},t),\end{equation}
where $[n_{2D}]=cm^{-2}$, $\textbf{r}_{2D}=\{x,y\}$ is the coordinate vector in the two dimensional physical space, $R_{2N}=\{\textbf{r}_{2D,1}, ..., \textbf{r}_{2D,i}, ..., \textbf{r}_{2D,N}\}$ set of all coordinates in 2N configurational space with $\textbf{r}_{2D,i}=\{x_{i},y_{i}\}$, and $dR_{2N}=\prod_{i=1}^{N}d\textbf{r}_{2D,i}$ is the element of configurational space.

Definition (\ref{Bscf_Fer def density 2D}) allows to derive the continuity equation and the Euler equation:
\begin{equation}\label{Bscf_Fer cont eq 2D}
\partial_{t}n_{2D}+\nabla\cdot(n_{2D}\textbf{v})=0,\end{equation}
and
$$mn_{2D}(\partial_{t}+\textbf{v}\cdot\nabla)\textbf{v}+\nabla p_{2D}-\frac{\hbar^{2}}{4m}n_{2D}\nabla\Biggl(\frac{\triangle n_{2D}}{n_{2D}}-\frac{(\nabla n_{2D})^{2}}{2n_{2D}^{2}}\Biggr)$$
$$=
\textrm{P}^{\beta}_{2D}\nabla \textrm{E}^{\beta}_{ext}+P^{\beta}_{2D}\nabla\int d\textbf{r}'_{2D}G^{\beta\gamma}(\textbf{r},\textbf{r}')P^{\gamma}_{2D}(\textbf{r}',t)$$
\begin{equation}\label{Bscf_Fer bal imp eq 2D} +\zeta_{2D}\sqrt{2\pi}d^{2}I_{2}\nabla n^{\frac{5}{2}},\end{equation}
where $I_{2}=8.045$, in this section, $\nabla=\nabla_{2D}=\textbf{i}\partial_{x}+\textbf{j}\partial_{y}$, $\triangle=\triangle_{2D}=\partial_{x}^{2}+\partial_{y}^{2}$, $\textbf{v}=\textbf{v}_{2D}=\{v_{x},v_{y}\}$, $\textbf{P}=\{P_{x}, P_{y}, P_{z}\}$, $P_{z}$ is the projection of the polarisation on the direction perpendicular to the plane, with $\beta=\gamma=z$ in equation (\ref{Bscf_Fer bal imp eq 2D}), $p_{2D}$ is the two dimensional pressure of degenerate partially spin polarised spin-1/2 fermions, its explicit form is
\begin{equation}\label{Bscf_Fer}p_{2D}=p_{2D\Updownarrow}= (1+\eta^{2})\pi\hbar^{2}n_{a,2D}^{2}/(2m_{a}),\end{equation}
parameter $\zeta_{2D}$ describes dependence of the exchange electric dipole-dipole interaction on the equilibrium spin condition:
\begin{equation}\label{Bscf_Fer} \zeta_{2D}=(1+\eta)^{5/2}-(1-\eta)^{5/2},\end{equation}
and
\begin{equation}\label{Bscf_Fer} G^{zz}=-\frac{1}{\xi^{3}}+\frac{1}{3}\triangle_{2D}\frac{1}{\xi}\end{equation}
is an explicit form of the zz matrix element of the Green function of the electric dipole interaction, where we have applied formula (\ref{Bscf_Fer dd int Green funct zz}) assuming that the two dimensional layer of fermions is located in plane $z=0$. Next we can use $\triangle_{2D}\frac{1}{\xi}=\frac{1}{\xi^{3}}$ to obtain the final form of $G^{zz}$ for two dimensional systems
\begin{equation}\label{Bscf_Fer dd int Green funct zz 2D short} G^{zz}=-\frac{2}{3}\frac{1}{\xi^{3}}.\end{equation}
Applying formula (\ref{Bscf_Fer dd int Green funct zz 2D short}) in the Euler equation (\ref{Bscf_Fer bal imp eq 2D}) we can calculate spectrum of the collective excitations of two dimensional dipolar fermions.

Three dimensional dipolar quantum gases have one preferable direction $\textbf{E}_{ext}$, while two dimensional dipolar quantum gases have two preferable directions $\textbf{E}_{ext}$ and $\textbf{n}$, where $\textbf{n}$ is the perpendicular to the plane. In this paper we do not use this degree of freedom and assume $\textbf{E}_{ext}\parallel\textbf{n}$. We focus our attention on the electric field perpendicular to the sample, so we have $\textbf{P}_{2D}=n_{2D}d\textbf{l}$, with $\textbf{l}\parallel \textbf{e}_{z}$.

The spectrum of collective excitations propagating as the plane waves in the two dimensional structure of dipolar ultracold fermions appears as follows
$$\omega^{2}=-\frac{4\pi d^{2}n_{0}k^{3}}{3m}- \zeta_{2D}\frac{5}{2}\sqrt{2\pi}I_{2}\frac{d^{2}n_{0,2D}^{3/2}k^{2}}{m}$$
\begin{equation}\label{Bscf_Fer spectrum 2D} +\vartheta_{2D}\frac{\pi\hbar^{2}n_{0,2D}}{m^{2}}k^{2}+\frac{\hbar^{2}k^{4}}{4m^{2}}.\end{equation}
As the three-dimensional and quasi two-dimensional spectrums, solution (\ref{Bscf_Fer spectrum 2D}) consists of four parts. The first terms describes the contribution of the self-consistent field part of electric dipole-dipole interaction under assumption that the external electric field creating the equilibrium polarisation is perpendicular to the sample. The second term presents the contribution of the exchange part of the electric dipole-dipole interaction. Both of them are negative. The third term is the Fermi pressure contribution derived for system of particles moving inside the plane. The last term is the quantum Bohm potential.

Let us note that we have considered systems of two-dimensional aligned dipolar fermions with the exchange interaction. The spectrum of collective excitations of dipolar fermions with the account of the dipole direction evolution, but with no account of the exchange dipole-dipole interaction, was obtained in Refs. \cite{Andreev LongetDipFermi12} for the three-dimensional systems, and \cite{Andreev RPJ 13} for the two-dimensional systems.

It is essential to note the differences between spectrum of the quasi two-dimensional and the two-dimensional dipolar fermions (\ref{Bscf_Fer spectrum quasi2D small k}) and (\ref{Bscf_Fer spectrum 2D}). The quasi two-dimensional regime (\ref{Bscf_Fer spectrum quasi2D small k}) contains trace of the three dimensional equation of state, being proportional to $n_{q2D,0}^{2/3}$, while the two-dimensional limit gives $\omega^{2}\sim n_{0,2D}$.

At the eddy-free motion $\textbf{v}_{2D}=\nabla_{2D}\phi$ of the two-dimensional medium of dipolar fermions with the aligned electric dipoles the hydrodynamic equations (\ref{Bscf_Fer cont eq 2D}), (\ref{Bscf_Fer bal imp eq 2D}) can be rewritten in the form of the non-linear Schrodinger equation
$$\imath\hbar\partial_{t}\Phi_{2D}(\textbf{r},t)=\Biggl(-\frac{\hbar^{2}}{2m}\triangle_{2D}+\vartheta_{2D}\frac{\pi\hbar^{2}}{m}n-\textbf{d}\textbf{E}_{ext}$$
\begin{equation}\label{Bscf_Fer nlse 2D with exchange} -d^{2}\int d\textbf{r}'_{2D}G_{2D}^{zz}(\textbf{r},\textbf{r}')n_{2D}(\textbf{r}',t)-\zeta_{2D}\frac{5}{3}\sqrt{2\pi}I_{2}d^{2}n^{\frac{3}{2}} \Biggr)\Phi_{2D}(\textbf{r},t).\end{equation}
The NLSE (\ref{Bscf_Fer nlse 2D with exchange}) is obtained for the effective macroscopic wave function $\Phi_{2D}(\textbf{r},t)$ defined in terms of hydrodynamic variables: $\Phi_{2D}(\textbf{r},t)=\sqrt{n_{2D}}\exp(\imath m\phi/\hbar)$.

\section{Discussions and Conclusions}

The quantum hydrodynamic method for ultracold dipolar fermions descriptions for the aligned dipoles has been developed. This method describes the electric dipole-dipole interaction as a long-range interaction. However, the self-consistent field approximation has not been applied in this paper. We have explicitly considered the two-particle hydrodynamic concentration appearing in the Euler equation in the force field of dipole-dipole interaction. We have considered this function  in the limit of weak interparticle interaction. At approximate calculation of the two-particle concentration we have explicitly considered antisymmetric N-particle wave function as the Slater determinant. The final result for the force field arises as the sum of two parts: the self-consistent field part and the exchange part. Analysis of the force field shows that the exchange interaction can reach same magnitude as the self-consistent field part. However the exchange part of the  force field strongly depend on distribution of fermions over quantum states.

Spin-1/2 fermions at temperatures considerably less than the Fermi temperature may occupy one of two spin states. In this case we can put one fermion in each quantum state in the momentum space, hence fermions occupy all states with the momentum smaller than $\sqrt[3]{2}\tilde{p}_{Fe}$, where $\tilde{p}_{Fe}\equiv \tilde{p}_{Fe,3D}=(3\pi^{2})^{\frac{1}{3}}n_{0}^{\frac{1}{3}}\hbar$ is the Fermi momentum, with the equilibrium particle concentration $n_{0}$ (or, in the case of two dimensional plane-like distribution of particles, they occupy all states with the momentum smaller than $\sqrt{2}\tilde{p}_{Fe,2D}$, where $\tilde{p}_{Fe,2D}=\sqrt{2\pi n_{0,2D}}\hbar$ is the 2D Fermi momentum). In this case we have the fully spin polarised systems of spin-1/2 fermions possessing \emph{the electric dipole moments}.

The opposite limit case, when pairs of fermions with the opposite spins occupy each quantum state in the momentum space. In this case fermions occupy all states with the momentum smaller than $p_{Fe,3D}$ (or $p_{Fe,2D}$ for plane like structure). In this case we have no spin polarisation. The exchange  interaction force monotonically depends on the spin polarisation. Therefore, the exchange force field equals to zero at the zero spin polarisation. While it has maximum value at the full spin polarisation.

Considering spin-1/2 atoms and molecules, when all of them located in the same state of the fine structure, we have rather large contribution of the exchange electric dipole-dipole interaction in the Euler equation. This contribution is the same order as the self-consistent field part. Since they have opposite sings (repulsing self-consistent field part and attractive exchange part), they can cancel each other at some conditions.


Spectrum of the collective excitations of ultracold electric dipolar fermions contains the positive anisotropic contribution of the self-consistent field part of the dipole-dipole interaction ($\sim\cos^{2}\theta$), and the isotropic negative contribution of the exchange part of the dipole-dipole interaction. Together they give the term proportional "$\cos^{2}\theta-2/3$". If the dipole-dipole interaction dominate over the short-range interaction, then the obtained spectrum shows the famous roton instability.

Usually authors obtain the roton instability in dipolar BECs and dipolar fermions in the self-consistent field approximation due to consideration of the reduced potential energy of the dipole-dipole interaction, instead of the full potential considered in this paper. Here, the roton instability arises at another angle (we have "$\cos^{2}\theta-2/3$" instead of "$\cos^{2}\theta-1/3$"), and it has another physical mechanism, i.e. the exchange part of the electric dipole-dipole interaction. We should note that similar approximation for the dipolar BECs does not lead to any instability (see Ref. \cite{Andreev Exchange_DD_BEC}).

\begin{acknowledgements}
The author wish to thank Professor L.S. Kuz'menkov for discussions of the results obtained.
\end{acknowledgements}


\begin{thebibliography}{17}

\bibitem{Ni Science 08} K.-K. Ni, S. Ospelkaus, M. H. G. de Miranda, A. Pe'er,
B. Neyenhuis, J. J. Zirbel, S. Kotochigova, P. S. Julienne, D. S. Jin,
J. Ye, Science \textbf{322}, 231 (2008).

\bibitem{Ospelkaus PRL 10}  S. Ospelkaus, K.-K. Ni, G. Quemener,  B. Neyenhuis, D. Wang,
 M. H. G.de Miranda, J. L. Bohn, J. Ye, D. S. Jin, Phys. Rev. Lett. \textbf{104}, 030402 (2010).

\bibitem{Ospelkaus NatP 08} S. Ospelkaus, A. Pe'er, K. Ni, J. J. Zirbel,
B. Neyenhuis, S. Kotochigova, P. S. Julienne, J. Ye, D. S. Jin, Nat. Phys. \textbf{4}, 622 (2008).

\bibitem{Ospelkaus FD 09} S. Ospelkaus, K.-K. Ni, M. H. G. de Miranda,
B. Neyenhuis, D. Wang, S. Kotochigova, P. S. Julienne, D. S. Jin, J. Ye, Faraday Discuss. \textbf{142}, 351 (2009).


\bibitem{Park arx 11} J. W. Park, C.-H. Wu, I. Santiago, T. G. Tiecke, P. Ahmadi,
M. W. Zwierlein, Phys. Rev. A 85, 051602(R) (2012).


\bibitem{Deh PRA 10} B. Deh, W. Gunton, B. G. Klappauf, Z. Li, M. Semczuk,
J. Van Dongen, K. W. Madison, Phys. Rev. A \textbf{82}, 020701 (2010).

\bibitem{Hara PRL 11} H. Hara, Y. Takasu, Y. Yamaoka, J. M. Doyle, Y. Takahashi, Phys. Rev. Lett. \textbf{106}, 205304 (2011).

\bibitem{Ivanov PRL 11} V. V. Ivanov, A. Khramov, A. H. Hansen, W. H. Dowd,
F. Munchow, A. O. Jamison, S. Gupta, Phys. Rev. Lett. \textbf{106}, 153201 (2011).

\bibitem{Hansen PRA 11} A. H. Hansen, A. Khramov, W. H. Dowd, A. O. Jamison,
V. V. Ivanov, S. Gupta, Phys. Rev. A \textbf{84}, 011606 (2011).

\bibitem{Heo PRA 12} M.-S. Heo, T. T. Wang, C. A. Christensen, T. M. Rvachov,
D. A. Cotta, J.-H. Choi, Y.-R. Lee,W. Ketterle, Phys. Rev. A \textbf{86}, 021602 (2012).



\bibitem{Deiglmayr PRL 08} J. Deiglmayr, A. Grochola, M. Repp, K. Mortlbauer, C. Gluck, J. Lange, O. Dulieu, R. Wester,
 M. Weidemuller, Phys. Rev. Lett. \textbf{101}, 133004 (2008).

\bibitem{Sage PRL 05} J. M. Sage, S. Sainis, T. Bergeman, D. DeMille, Phys. Rev. Lett. \textbf{94}, 203001 (2005).

\bibitem{Debatin PCCP 11}  M. Debatin, T. Takekoshi, R. Rameshan,
 L. Reichsollner, F. Ferlaino, R. Grimm, R. Vexiau, N. Bouloufa, O. Dulieu, H.-C. Nagerl, Phys. Chem. Chem. Phys. \textbf{13}, 18926 (2011).

\bibitem{Lercher EPJD 11} A. Lercher, T. Takekoshi, M. Debatin, B. Schuster, R. Rameshan,
F. Ferlaino, R. Grimm, H. Nagerl, Eur. Phys. J. D \textbf{65}, 3 (2011).

\bibitem{Takekoshi PRA 12} T. Takekoshi, M. Debatin, R. Rameshan, F. Ferlaino, R. Grimm,
H.-C. Nagerl, C. R. Le Sueur, J. M. Hutson, P. S. Julienne, S. Kotochigova, E. Tiemann, Phys. Rev. A \textbf{85}, 032506 (2012).

\bibitem{Aikawa PRL 10} K. Aikawa, D. Akamatsu, M. Hayashi, K. Oasa, J. Kobayashi,
P. Naidon, T. Kishimoto, M. Ueda, S. Inouye, Phys. Rev. Lett. \textbf{105}, 203001 (2010).

\bibitem{Kerman PRL 04} A. J. Kerman, J. M. Sage, S. Sainis, T. Bergeman,
D. DeMille, Phys. Rev. Lett. \textbf{92}, 033004 (2004).

\bibitem{Mancini PRL 04} M. W. Mancini, G. D. Telles, A. R. L. Caires, V. S. Bagnato,
L. G. Marcassa, Phys. Rev. Lett. \textbf{92}, 133203 (2004).

\bibitem{Haimberger PRA 04} C. Haimberger, J. Kleinert, M. Bhattacharya, N. P. Bigelow, Phys. Rev. A \textbf{70}, 021402 (2004).

\bibitem{Haimberger NJP 09} C. Haimberger, J. Kleinert, P. Zabawa, A. Wakim, N. P. Bigelow, New J. Phys. \textbf{11}, 055042 (2009).


\bibitem{Shuman PRL 09} E. S. Shuman, J. F. Barry, D. R. Glenn, D. DeMille, Phys. Rev. Lett. \textbf{103}, 223001 (2009).
\bibitem{Shuman Nat 10} E. S. Shuman, J. F. Barry, D. DeMille, Nature \textbf{467}, 820 (2010).


\bibitem{Zabawa PRA 10}  P. Zabawa, A. Wakim, A. Neukirch, C. Haimberger, N. P. Bigelow,
A. V. Stolyarov, E. A. Pazyuk, M. Tamanis, R. Ferber, Phys. Rev. A , \textbf{82}, 040501 (2010).

\bibitem{Wang PRL 04} D. Wang, J. Qi, M. F. Stone, O. Nikolayeva, H. Wang, B. Hattaway,
S. D. Gensemer, P. L. Gould, E. E. Eyler, W. C. Stwalley, Phys. Rev. Lett. \textbf{93}, 243005 (2004).

\bibitem{Nemitz PRA 09} N. Nemitz, F. Baumer, F. Munchow, S. Tassy, A. Gorlitz, Phys. Rev. A \textbf{79}, 061403 (2009).

\bibitem{Gabbanini PCCP 11} C. Gabbanini, O. Dulieu, Phys. Chem. Chem. Phys., \textbf{13}, 18905 (2011).

\bibitem{Ridinger EPL 11} A. Ridinger, S. Chaudhuri, T. Salez,
D. R. Fernandes, N. Bouloufa, O. Dulieu, C. Salomon, F. Chevy, Europhys. Lett. \textbf{96}, 33001 (2011).

\bibitem{Ji PRA 12} Z. Ji, H. Zhang, J. Wu, J. Yuan, Y. Yang, Y. Zhao, J. Ma,
L. Wang, L. Xiao, S. Jia, Phys. Rev. A \textbf{85}, 013401 (2012).

\bibitem{Cho EPJD 11} H. Cho, D. McCarron, D. Jenkin, M. KAuppinger, S. Cornish, Eur. Phys. J. D \textbf{65}, 125 (2011).

\bibitem{McCarron PRA 11} D. J. McCarron, H. W. Cho, D. L. Jenkin, M. P. Koppinger, S. L. Cornish, Phys. Rev. A, \textbf{84}, 011603 (2011).

\bibitem{Munchow PCCP 11} F. Munchow, C. Bruni, M. Madalinski, A. Gorlitz, Phys. Chem. Chem. Phys. \textbf{13}, 18734 (2011).





\bibitem{Lahaye RPP 09} T. Lahaye, C. Menotti, L. Santos, M. Lewenstein, and T. Pfau, Rep. Prog. Phys. \textbf{72}, 126401 (2009).
\bibitem{Lewenstein AinP 07} M. Lewenstein, A. Sanpera, V. Ahufinger,
B. Damski, A. Sen(De), and U. Sen, Advances in Physics, \textbf{56}, 243 (2007).
\bibitem{Carr NJP 09} L. D. Carr et al., New J. Phys. \textbf{11}, 055049 (2009).
\bibitem{Baranov P R 08} M. A. Baranov, Physics Reports \textbf{464}, 71 (2008).
\bibitem{Kawaguchi Ph Rep 12} Yuki Kawaguchi, Masahito Ueda, Physics Reports \textbf{520},  253 (2012).
\bibitem{Quemener CR 12} Goulven Quemener, and Paul S. Julienne, Chem. Rev. \textbf{112}, 4949 (2012).
\bibitem{Baranov CR 12} M. A. Baranov, M. Dalmonte, G. Pupillo, and P. Zoller, Chem. Rev.  \textbf{112},  5012 (2012).


\bibitem{Andreev EPJ D 13} P. A. Andreev and L. S. Kuz'menkov, Eur. Phys. J. D \textbf{67}, 216 (2013). 
\bibitem{Andreev RPJ 12} P. A. Andreev, Russian Physics Journal \textbf{54}, 1360 (2012).



\bibitem{Andreev JP B 14} P. A. Andreev, L. S. Kuz'menkov, Journal of Physics B: Atomic, Molecular and Optical Physics \textbf{47}, 225301 (2014).

\bibitem{Andreev RPJ 13} P. A. Andreev, Russian Physics Journal \textbf{55}, N. 10, p. 1190 (2013).




\bibitem{Giorgini RMP 08} S. Giorgini, L. P. Pitaevskii, S. Stringari, Rev. Mod. Phys. \textbf{80}, 1215 (2008).




\bibitem{Belemuk PRA 07} A. M. Belemuk, V. N. Ryzhov, and S.-T.
Chui, Phys. Rev. A \textbf{76}, 013609(2007).

\bibitem{Adhikari PRA 04}S. K. Adhikari, Phys. Rev. A \textbf{70},
043617 (2004).

\bibitem{Adhikari PRA05} S. K. Adhikari, Phys. Rev. A \textbf{72}, 053608 (2005).

\bibitem{Adhikari JPB05} S. K. Adhikari, Journal of Physics B \textbf{38}, 3607 (2005).

\bibitem{Adhikari PRA 07} S. K. Adhikari, and Luka
Salasnich, Phys. Rev. A \textbf{75}, 053603(2007).

\bibitem{Adhikari NJP06} S. K. Adhikari, New Journal of Physics \textbf{8}, 258 (2006).

\bibitem{Bludov PRA06} Yu. V. Bludov, J. Santhanam, V. M. Kenkre, and V. V. Konotop, Phys. Rev. A \textbf{74}, 043620 (2006).

\bibitem{Rizzi PRA08} Matteo Rizzi and Adilet Imambekov, Phys. Rev. A \textbf{77}, 023621 (2008).

\bibitem{Maruyama PRA08} Tomoyuki Maruyama and George F. Bertsch, Phys. Rev. A \textbf{77}, 063611 (2008).

\bibitem{Karpiuk PRA06} Tomasz Karpiuk, Mirosaw Brewczyk and Kazimierz Rzewski, Phys. Rev. A \textbf{73}, 053602 (2006).






\bibitem{Lima PRA 10} A. R. P. Lima and A. Pelster, Phys. Rev. A \textbf{81}, 063629 (2010).




\bibitem{Santos PRL 00} L. Santos, G.V. Shlyapnikov, P. Zoller, and M.
Lewenstein, Phys. Rev. Lett. \textbf{85}, 1791 (2000).

\bibitem{Yi PRA 00} S. Yi and L. You, Phys. Rev. A, \textbf{61}, 041604(R) (2000).

\bibitem{Goral PRA 00} K. Goral, K. Rzazewski, and T.
Pfau, Phys. Rev. A \textbf{61}, 051601(R) (2000).





\bibitem{Yi PRA 01} S. Yi and L. You, Phys. Rev. A, \textbf{63}, 053607 (2001).


\bibitem{Babadi_Demler PRA 12} Mehrtash Babadi and Eugene Demler, Phys. Rev. A \textbf{86}, 063638 (2012).


\bibitem{Lima PRA 12} A. R. P. Lima, and A. Pelster, Phys. Rev. A \textbf{86}, 063609 (2012).

\bibitem{Wang NJP 08} Daw-Wei Wang, New Journal of Physics \textbf{10},  053005
(2008).

\bibitem{Ticknor PRA 12} Christopher Ticknor, Phys. Rev. A \textbf{85}, 033629 (2012).

\bibitem{Natu_Wilson PRA 13} Stefan S. Natu, Ryan M. Wilson, Phys. Rev. A \textbf{88}, 063638 (2013).




\bibitem{Baillie PRA 12} D. Baillie and P. B. Blakie, Phys. Rev. A \textbf{86}, 023605 (2012).

\bibitem{Baillie PRA 12 (R)} D. Baillie and P. B. Blakie, Phys. Rev. A \textbf{86}, 041603(R) (2012).


\bibitem{Andreev MPL 13} P. A. Andreev, Mod. Phys. Lett. B \textbf{27}, 1350096 (2013).

\bibitem{Andreev EPJ D 14}	 P. A. Andreev, L. S. Kuz'menkov, Eur. Phys. J. D \textbf{68}, 270 (2014).


\bibitem{Andreev LongetDipFermi12} P. A. Andreev, arXiv:1209.4196.



\bibitem{Fischer PRA 06R} Uwe R. Fischer, Phys. Rev. A \textbf{73}, 031602(R) (2006).
\bibitem{Goral PRA 02} K. Goral and L. Santos, Phys. Rev. A \textbf{66}, 023613
(2002).
\bibitem{Santos PRL 03} L. Santos, G. V. Shlyapnikov, and M. Lewenstein, Phys. Rev.
Lett. \textbf{90}, 250403 (2003).
\bibitem{Dell PRL 03} D. H. J. O'Dell, S. Giovanazzi, and G. Kurizki, Phys. Rev.
Lett. \textbf{90}, 110402 (2003).
\bibitem{Giovanazzi EPJD 04} S. Giovanazzi and D. H. J. O'Dell, Eur. Phys. J. D \textbf{31}, 439
(2004).
\bibitem{Ronen PRL 07} S. Ronen, D. C. E. Bortolotti, and J. L. Bohn, Phys. Rev.
Lett. \textbf{98}, 030406 (2007).
\bibitem{Wilson PRL 08} R. M. Wilson, S. Ronen, J. L. Bohn, and H. Pu, Phys.
Rev. Lett. \textbf{100}, 245302 (2008).


\bibitem{Baillie_arx 14} D. Baillie and P. B. Blakie, arXiv: 1407.4252v1.

\bibitem{Natu+Sarma PRA 14} Stefan S. Natu, L. Campanello, and S. Das Sarma, Phys. Rev. A \textbf{90}, 043617 (2014).

\bibitem{Ticknor PRL 11} C. Ticknor, R. M. Wilson, and J. L. Bohn, Phys. Rev. Lett. \textbf{106}, 065301 (2011).
\bibitem{Fedorov PRA 14} A. K. Fedorov, I. L. Kurbakov, Y. E. Shchadilova, and Yu. E. Lozovik, Phys. Rev. A \textbf{90}, 043616 (2014).

\bibitem{Sieberer ar 11} L. M. Sieberer, and M. A.Baranov, Phys. Rev. A \textbf{84}, 063633 (2011).



\bibitem{Babadi arx 12} M. Babadi, B. Skinner, M. M. Fogler, E. Demler, Europhysics Letters \textbf{103}, 16002 (2013). 

\bibitem{Marchetti arx 12} F. M. Marchetti, and M. M. Parish, Phys. Rev. B \textbf{87}, 045110 (2013). 
\bibitem{Block Bruun arx 14} J. K. Block, and G. M. Bruun, Phys. Rev. B \textbf{90}, 155102 (2014).

\bibitem{Baranov arx 10} M. A. Baranov, A. Micheli, S. Ronen, and P. Zoller, Phys. Rev. A \textbf{83}, 043602 (2011). 

\bibitem{Liu Li Yin} Bo Liu, Xiaopeng Li, Lan Yin, and W. Vincent Liu, arXiv:1407.2949.

\bibitem{Adhikari ar 12 fermi} S. K. Adhikari, J. Phys B \textbf{45}, 235303 (2012). 
\bibitem{Wilson PRA 12} R. M. Wilson, C. Ticknor, J. L. Bohn, and E. Timmermans,
Phys. Rev. A \textbf{86}, 033606 (2012).
\bibitem{Adhikari arx 13 mixt} S. K. Adhikari, Phys. Rev. A \textbf{88}, 043603  (2013). 

\bibitem{Jona-Lasinio 13} M. Jona-Lasinio, K.  Lakomy and L. Santos,  Phys. Rev. A \textbf{88}, 013619 (2013). 

\bibitem{Macia PRA 14} A. Macia, J. Boronat, and F. Mazzanti, Phys. Rev. A \textbf{90}, 061601(R) (2014).

\bibitem{Kotochigova arx 14} Svetlana Kotochigova, Rep. Prog. Phys. \textbf{77}, (2014). 093901 





\bibitem{MaksimovTMP 2001 b} L. S. Kuz'menkov, S. G. Maksimov, and V. V. Fedoseev, Theor.
Math. Fiz. \textbf{126} 258 (2001) [Theoretical and Mathematical
Physics, \textbf{126} 212 (2001)].


\bibitem{Andreev Exchange_DD_BEC} P. A. Andreev, arXiv:1410.8716.


\bibitem{MaksimovTMP 1999} L. S. Kuz'menkov and S. G. Maksimov, Theoretical and Mathematical Physics \textbf{118} 227 (1999).
\bibitem{MaksimovTMP 2001} L. S. Kuz'menkov, S. G. Maksimov, and V. V. Fedoseev, Theoretical and Mathematical
Physics, \textbf{126} 110 (2001).

\bibitem{Andreev PRA08} P. A. Andreev, L. S. Kuz'menkov, Phys. Rev. A \textbf{78}, 053624
(2008).

\bibitem{Andreev AoP 14} P. A. Andreev, Annals of Physics, \textbf{350}, 198 (2014).

\bibitem{Andreev 1407 Review} P. A. Andreev, L. S. Kuz'menkov, arXiv:1407.7770.

\bibitem{Nozieres PR 58} P. Nozieres and D. Pines, Phys. Rev. \textbf{111}, 442 (1958).
\bibitem{Kanazawa PTP 60} H. Kanazawa, S. Misawa and K. Fujita, Progr. Theoret. Phys. (Kyoto) \textbf{23}, 426 (1960).
\bibitem{DuBois AnnP 59} D. F. DuBois, Ann. of Phys. \textbf{8}, 24 (1959).
\bibitem{Hedin JP C 71} L. Hedin, and B. I. Lundqvist,  J. Phys. C: Solid St. Phys., \textbf{4}, 2064 (1971).
\bibitem{Brey PRB 90} L. Brey, Jed Dempsey, N. F. Johnson, and B. I. Halperin, Phys. Rev. B \textbf{42}, 1240 (1990).
\bibitem{Datta JAP 83} S. Datta and R. L. Gunshor, Journal of Applied Physics \textbf{54}, 4453 (1983). 

\bibitem{Zyl arXiv14} B. P. van Zyl, E. Zaremba, and J. Towers, Phys. Rev. A \textbf{90}, 043621 (2014). 


\bibitem{Andreev PRB 11} P. A. Andreev, L. S. Kuzmenkov, M. I. Trukhanova, Phys. Rev. B \textbf{84}, 245401 (2011).

\bibitem{Andreev IJMP B 13} P. A. Andreev, Int. J. Mod. Phys. B \textbf{27}, 1350017 (2013).


\bibitem{Petrov PRL 00} D. S. Petrov, M. Holzmann, and G. V. Shlyapnikov, Phys. Rev.
Lett. \textbf{84}, 2551 (2000).

\bibitem{Lu_Shlyapnikov arX_14} Zhen-Kai Lu, D. S. Petrov, G. V. Shlyapnikov,  arXiv:1409.7737.
\bibitem{Boudjemaa PRA 13} A. Boudjemaa and G. V. Shlyapnikov, Phys. Rev. A \textbf{87}, 025601 (2013).

\bibitem{Fetter AoP 73} A. L. Fetter, Ann. Phys. \textbf{81}, 367 (1973).
\bibitem{Fetter AoP 74} A. L. Fetter, Ann. Phys. \textbf{88}, 1 (1974).
\bibitem{Oji PRB 86} H. C. A. Oji and A. H. MacDonald, Phys. Rev. B \textbf{33}, 3810 (1986).
\bibitem{Batke PRB 86} E. Batke, D. Heitmann, and C.W. Tu, Phys. Rev. B \textbf{34}, 6951 (1986).

%
\end{thebibliography}
\end{document}